\documentclass[a4paper, 10 pt, conference]{ieeeconf}
\usepackage[utf8]{inputenc}
\usepackage{amsmath}
\usepackage{amssymb}
\usepackage{graphicx}
\usepackage{caption}
\usepackage{babel}
\usepackage{balance}
\usepackage{cite}
\usepackage[position=bottom]{subfig}
\usepackage{floatrow}
\usepackage{xcolor}

\IEEEoverridecommandlockouts

\let\theoremstyle\relax
\usepackage{amsthm}

\usepackage{float}
\floatstyle{plaintop}
\restylefloat{table}

\theoremstyle{definition}

\newtheorem{assumption}{Assumption}

\usepackage{comment}
\usepackage{todonotes}

\usepackage[ruled]{algorithm2e}

\title{Conference2022}
\title{\LARGE \bf Performance Analysis of Optimally Coordinated Connected and Automated Vehicles in a Mixed Traffic Environment}

\author{Alejandra Valencia, {\itshape{Student Member, IEEE}}, A M Ishtiaque Mahbub, {\itshape{Student Member, IEEE}},\\ Andreas A. Malikopoulos, {\itshape{Senior Member, IEEE}}
\thanks{This research was supported by ARPAE's NEXTCAR program under the award number DE-AR0000796.}
\thanks{The authors are with the Department of Mechanical Engineering, University of Delaware, Newark, DE 19716 USA (emails: \tt\small{aleval@udel.edu}; \tt\small{mahbub@udel.edu}; \tt\small{andreas@udel.edu}.)}}

\date{January 2022}

\begin{document}

\maketitle

\begin{abstract}
Trajectory planning of connected and automated vehicles (CAVs) poses significant challenges in a mixed traffic environment due to the presence of human-driven vehicles (HDVs). In this paper, we apply a framework that allows coordination of CAVs and HDVs traveling through a traffic corridor consisting of an on-ramp merging, a speed reduction zone, and a roundabout. We study the impact of different penetration rates of CAVs and traffic volumes on the efficiency of the corridor. We provide extensive simulation results and report on the benefits in terms of total travel time and fuel economy.
\end{abstract}

\section{Introduction}

The emergence of connected and automated vehicles (CAVs) enables a novel computational framework that can provide optimal control actions in real time. Researchers have demonstrated that optimal control and coordination of CAVs can alleviate congestion at different traffic scenarios, reduce emission, improve fuel efficiency and increase passenger safety \cite{Margiotta2011, chalaki2020TITS, letter2017merging}.
%
%
Several efforts in the literature have addressed the problem of optimal coordination of CAVs to improve the vehicle, and network-level performances \cite{mosebach2016merging, Guanetti2018, DeLaFortelle2010, Dresner2008}. Recent efforts have reported results on coordination of CAVs at on-ramp merging roadways, roundabouts, speed reduction zones, signal-free intersections, and traffic corridors (see \cite{Ntousakis:2016aa,Zhao2018ITSC, mahbub2020decentralized, mahbub2020ACC-2, kotsialos2004Merging,Kumaravel:2021wi}). 

It is expected that CAVs will gradually penetrate the market and interact with human-driven vehicles (HDVs) by 2060 \cite{alessandrini2015automatedmixed2060}. However, different penetration rates of CAVs can significantly alter transportation efficiency and safety. While the aforementioned studies have shown the benefits of CAVs to reduce energy consumption and alleviate traffic congestion in specific traffic scenarios, most of these efforts have focused on 100\% CAV penetration rates without considering the HDVs. 

One of the research directions towards controlling the CAVs in a mixed traffic environment has been the development of adaptive cruise control \cite{zheng2017platooning, sharon2017protocol, milanes2013cooperative} where a CAV, preceded by a single or a group of HDVs, applies cruise control to optimize a given objective, e.g., improvement of fuel economy \cite{jin2017fuel_mixplatoon}, minimization of backward propagating wave \cite{hajdu2019robust, orosz2016connected}, etc.  
Although these research efforts \cite{guler2014_Intersection} aim at enhancing our understanding of improving the efficiency through coordination of CAVs in a mixed traffic environment, deriving a tractable solution to the problem of CAV coordination at merging or roundabout scenario still remains challenging. Several approaches reported in the literature implemented well-known car-following models, which emulate the human-driving behavior \cite{Gipps1981, wiedemann1974}, to derive a deterministic quantification of the vehicle trajectory \cite{zhang2018penetration, ding2020penetration}. Other approaches have used car-following models \cite{wan2016optimalmixed} or learning algorithms \cite{kreidieh2018dissipating, wu2017framework} for CAVs coordination in a mixed traffic environment. 

In this paper, we analyze the impact of optimally coordinating CAVs traveling through a mixed traffic corridor including three different scenarios: on-ramp merging, speed reduction zone, and roundabout. In this context, CAVs interact with HDVs at varying penetration rates and different traffic volumes.
The contributions of this paper are the (i) development of a simulation environment of an optimal CAV coordination framework at a corridor in a mixed traffic network, and (ii) a detailed analysis of the impact of CAV penetration on the vehicle, and network-level performance, in terms of fuel economy and travel time, under different traffic volumes.  

The remainder of the paper proceeds as follows. In Section \ref{sec:ModelingFramework}, we provide the modeling framework for a mixed traffic environment. In Section \ref{sec:control_framework}, we present the coordination framework for CAVs traveling through the traffic corridor while interacting with HDVs. In Section \ref{sec:Simulation}, we provide a detailed analysis of the simulation results. Finally, we draw concluding remarks in Section \ref{sec:conclusion}.

\section{Problem Formulation} \label{sec:ModelingFramework}
\begin{figure}[ht!]
    \centering
    \includegraphics[width=0.6\linewidth]{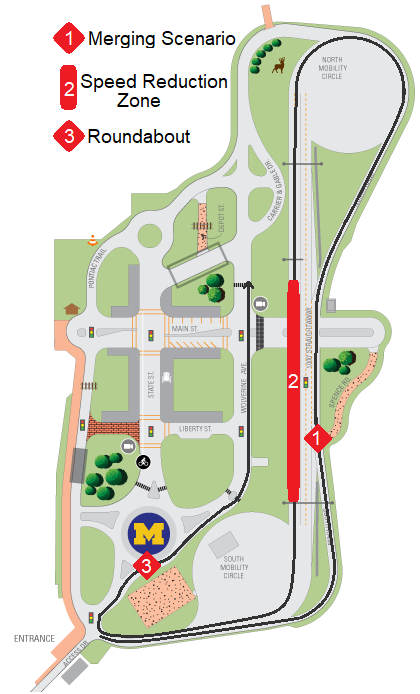}
    \caption{Corridor of Mcity with three traffic scenarios.}
    \label{fig:mcity_corridor}
\end{figure}

We consider the University of Michigan's Mcity where CAVs and HDVs are traveling through a particular test route as illustrated by the black trajectory in Fig. \ref{fig:mcity_corridor}. The route consists of three traffic scenarios, indexed by $z=1,2,3$, representing a highway on-ramp, a speed reduction zone, and a roundabout, respectively. To create traffic congestion in the test route, we consider additional traffic flow at the adjacent roads. 
%
Upstream of each traffic scenario, we define a \emph{control zone} where CAVs coordinate with each other to avoid any rear-end or lateral collision. The length of the control zone is $L_z\in\mathbb{R}^+$ for each traffic scenario $z$.
Since the HDVs do not share their state information, we consider the presence of \emph{coordinators}, which can be loop-detectors, roadside units, or comparable sensory devices, that collect the state information of the HDVs traveling within each control zone. The coordinators transmit the HDVs state information to each CAV within each control zone using standard vehicle-to-infrastructure communication protocol. Note that, the coordinators do not make any control decisions for the CAVs.

We define the area of potential lateral collision to be the \emph{merging zone} of length $S_z$ specific to traffic scenario $z$, as illustrated by the red marked area with numbers $1$, $2$, and $3$ in Fig. \ref{fig:mcity_corridor}. The objective of each CAV is to derive its optimal control input (acceleration/deceleration) to cross the traffic scenarios without any collision with the other CAVs and HDVs.
%
Let $t_i^{0,z}$ be the time when each vehicle $i$ enters the control zone towards traffic scenario $z$, and $t_i^{m,z}$ and $t_i^{f,z}$ be the times when each vehicle enters and exits the merging zone of the traffic scenario $z$, respectively. Let $\mathcal{N}_z=\{1,\dots,N(t)\}$, $ t\in \mathbb{R}^+$, be a queue of vehicles to be analyzed for traffic scenario $z$, where $N(t)$ is the total number of CAVs within the control zone of the specific traffic scenario $z$ at time $t \in \mathbb{R}^+$. We denote $\mathcal{N}_\text{cav}$ and $\mathcal{N}_\text{hdv}$ to be the sets of CAVs and HDVs such that $\mathcal{N}_\text{cav}\cup \mathcal{N}_\text{hdv} = \mathcal{N}_z$. 

The dynamics of each vehicle $i \in \mathcal{N}_z$ is modeled as a double integrator
\begin{equation} \label{eq:state}
\dot{p}_i = v_i(t), \quad \dot{v}_i = u_i(t),
\end{equation}
where $p_i(t) \in \mathcal{P}_i, v_i(t) \in \mathcal{V}_i$, and $u_i(t) \in \mathcal{U}_i$ denote the position, speed, and acceleration/deceleration (control input) of each vehicle $i$, respectively. The sets $\mathcal{P}_i,\mathcal{V}_i$, and $\mathcal{U}_i$ are complete and totally bounded subsets of $\mathbb{R}$. Let $\mathbf{x}_i(t)=[p_i(t)~ v_i(t)]^T$ denote the state of each vehicle $i$, with initial value $\mathbf{x}_i(t_i^{0,z})=[p_i(t_i^{0,z}) ~ v_i(t_i^{0,z})]^T$.

We impose the following state and control constraints in our modeling framework,
\begin{equation}\label{eq:constraints}
\begin{aligned} 
u_\text{min} &\leq u_i(t) \leq u_\text{max},~ \text{and} \\
0 &\leq v_\text{min} \leq v_i(t) \leq v_\text{max}, ~ \forall t \in [t_i^{0,z}, ~ t_i^{f,z}],
\end{aligned} 
\end{equation}
where $u_\text{min}, u_\text{max}$ are the minimum deceleration and maximum acceleration, $v_\text{min}, v_\text{max}$ are the minimum and maximum speed limits respectively.
Next, we consider the rear-end and lateral safety constraints as
\begin{equation} \label{eq:safety}
s_i(t)=p_k(t)-p_i(t)\geq \delta_i(t), ~\forall t \in  [t_i^{0,z}, ~ t_i^{f,z}],
\end{equation}
where vehicle $k$ is immediately ahead of $i$ on the same lane. 
Lateral collision between any two vehicles $i,j\in \mathcal{N}_z$ can be avoided if the following constraint holds
\begin{equation}
\Gamma_i \cap \Gamma_j=\varnothing, ~  \forall t\in [t_i^{m,z}, t_i^{f,z}], \quad i,j\in \mathcal{N}_z(t),
\label{eq:lateral_constraint}
\end{equation} 
where $\Gamma_i := \{t\,\,|\,t\in  [t_i^{m,z}, t_i^{f,z}]\}$.

For the CAV $i\in\mathcal{N}_\text{cav}$, the control input $u_i(t)$ in \eqref{eq:state} can be derived within the control zone, the structure of which we discuss in Section \ref{sec:control_framework}.
%
In contrast, we consider the Wiedemann car-following model proposed in \cite{wiedemann1974} to derive the control action of each HDV $i\in \mathcal{N}_{{hdv}}$.


In the modeling framework presented above, we impose the following assumptions.
\begin{assumption}\label{assumption:communication}
The communication among the CAVs and the coordinator occurs without any transmission latency, errors or data loss. 
\end{assumption}

\begin{assumption}\label{assumption:lane}
No lane change is allowed inside the control zone.
\end{assumption}

Assumption \ref{assumption:communication} might be strong, but can be relaxed as long as the noise in the measurements and/or delays is bounded. Assumption \ref{assumption:lane} simplifies the formulation by restricting the traffic flow to a single lane within the control zone. 

\section{Optimal Coordination Framework} \label{sec:control_framework}
For each CAV $i \in \mathcal{N}_\text{cav}$, we adopt the optimal control problem presented in \cite{mahbub2020sae-1}, i.e.,
\begin{gather} \label{eq:min}
\min_{u_i}\frac{1}{2}\int_{t_i^{0,z}}^{t_i^{m,z}} u_i^2(t)dt, ~\forall i \in \mathcal{N}_z, ~\forall z =1,2,3, \\
\text{subject to}: (\ref{eq:state}), (\ref{eq:constraints}),\nonumber\\
p_{i}(t_i^{0,z})=p_{i}^{0,z}\text{, }v_{i}(t_i^{0,z})=v_{i}^{0,z}\text{, }p_{i}(t_i^{m,z})=p_{z},\nonumber\\
\text{and given }t_i^{0,z}\text{, }t_i^{m,z},\nonumber
\end{gather}
where $p_{z}$ is the location (i.e., entry position) of merging zone $z$; $p_{i}^{0,z}$, $v_{i}^{0,z}$ are the initial position and speed of vehicle $i$ when it enters the control zone of traffic scenario $z$, respectively. The merging time $t_i^{m,z}$ can be obtained by solving an upper-level control problem including the safety constraints \eqref{eq:safety}, \eqref{eq:lateral_constraint} in an iterative manner, as detailed in \cite{mahbub2020sae-1}.
Suppose that, each CAV $i\in\mathcal{N}_z$ is aware of the information of the sets $\mathcal{L}_i^z$ and $\mathcal{C}_i^z$, which contain the unique id of the preceding vehicles traveling on the same lane and on a conflict lane relative to CAV $i$, respectively. Then, each CAV $i$ determines the time $t_i^{m,z}$ as follows (see \cite{mahbub2020sae-1}).
If vehicle $(i-1)\in \mathcal{L}^z_i$ and $(i-1)\in \mathcal{N}_\text{cav}$, we have
\begin{align}
    t_i^{m,z} = \max\bigg \{&\min\bigg\{t_{i-1}^{m,z}+\frac{\delta(v_i(t))}{v_{i-1}(t_{i-1}^{m,z})}, t_i^{0,z}+\frac{L_z}{v_{\min}}\bigg\},\nonumber\\&t_i^{0,z}+ \frac{L_z}{v_0(t_i^{0,z})}, t_i^{0,z}+\frac{L_z}{v_{\max}}\bigg\}, \label{eq:7}
\end{align}
while, if vehicle $(i-1)\in \mathcal{C}^z_i$ and $(i-1)\in \mathcal{N}_\text{cav}$, then 
\begin{align}
    t_i^{m,z} = \max\bigg \{&\min\bigg\{t_{i-1}^{m,z}+\frac{S_z}{v_{i-1}(t_{i-1}^{m,z})}, t_i^{0,z}+\frac{L_z}{v_{\min}}\bigg \}, \nonumber\\&t_i^{0,z}+\frac{L_z}{v_0(t_i^{0,z})}, t_i^{0,z}+\frac{L_z}{v_{\max}}\bigg \}, \label{eq:8}
\end{align}
where ${L_z}$ and ${S_z}$ are the length of the control zone and the length of the area of potential lateral collision, respectively.

Note that, if the vehicle preceding CAV $i$ is an HDV, i.e., $(i-1)\in \mathcal{N}_\text{hdv}$, then we apply $t_{i-1}^{m,z} = t_{i-1}^{0,z}+ \frac{L_z}{v_{i-1}( t_{i-1}^{0,z})}$ to estimate the merging time $t_{i-1}^{m,z}$ of HDV $i-1$. We then use  $t_{i-1}^{m,z}$ in \eqref{eq:7}-\eqref{eq:8} to derive the merging time $t_{i}^{m,z}$ of CAV $i$. The recursion of the above computation is initialized when the first vehicle enters the control zone.

Using Hamiltonian analysis \cite{bryson1975applied}, the unconstrained optimal control input $u^*_i(t)$ of CAV $i\in\mathcal{N}_z$ and the corresponding state trajectories at time $t \in [t_i^{0,z}, ~ t_i^{m,z}]$ are \cite{mahbub2020Automatica-2}
\begin{align}
&u^{*}_{i}(t) = a_{i} \cdot t + b_{i}, \label{eq:20}\\
&v^{*}_{i}(t) = \frac{1}{2} a_{i} \cdot t^2 + b_{i} \cdot t + c_{i},\label{eq:21}\\
&p^{*}_{i}(t) = \frac{1}{6}  a_{i} \cdot t^3 +\frac{1}{2} b_{i} \cdot t^2 + c_{i}\cdot t + d_{i}, \label{eq:22}%
\end{align}
where $a_i$, $b_{i}$, $c_{i}$, and $d_{i}$ are the constants of integration and can be computed using the analysis presented in \cite{mahbub2020Automatica-2}.

For the control of CAVs in a mixed environment, if the physically leading vehicle of a CAV is HDV, the CAV will probe the safety constraint continuously to make an adjustment to its motion primitive. A switching mechanism is applied in the study: the control algorithm for a CAV would always be switched on until the safety constraint \eqref{eq:safety} is activated in terms of the distance between itself and its preceding HDV.

\section{Simulation and Discussion} \label{sec:Simulation}

\subsection{Simulation Setup}
To implement the control framework presented in the previous section, we use the microscopic multi-modal commercial traffic simulation software PTV VISSIM \cite{ptv2014ptv, Vissim_microscopic} by creating a simulation environment replicating Mcity, as shown in Fig. \ref{fig:mcity_corridor}. The corridor through which the vehicle travels has a length of 1,200 m within the Mcity. The maximum and minimum acceleration values considered for each vehicle are 1.5 m/s$^2$ and -3.0 m/s$^2$, respectively. The speed limit on the on-ramp merging, speed reduction zone, and roundabout are 40 m/s, 18.6 m/s, and 25 m/s, respectively. The control zone length is 150 m and the safe headway time considered is 1.2 s. In our simulation study, we consider high, medium, and low traffic volumes for the test route as $500$, $400$ and $300$ vehicles per hour, and for the adjacent roads as $800$, $600$ and $400$ vehicles per hour, respectively. 
We consider the following three different cases:

\textbf{Baseline:} We construct it by considering all the vehicles to be HDVs and without any communication capability. The vehicles subscribe to the VISSIM built-in Wiedemann car following model \cite{wiedemann1974} to emulate the driving behavior of real human-driven vehicles. We adopt priority-based (yield/stop) traffic movement at the roundabout and on-ramp merging scenarios. 

\textbf{Optimal Coordination:} In this case, all the vehicles are CAVs, and communicate with each other inside the control zone. Therefore, they can optimize their travel time and fuel efficiency, and plan their optimal trajectories. We consider three isolated coordinators for each traffic scenario. For the uncontrolled paths in-between the control zones, the CAVs revert to the Wiedemann car following model \cite{wiedemann1974} to traverse their respective routes. To apply the optimal control framework, we override VISSIM's built-in car following module and associated attributes using the DriverModel API. 

\textbf{Partial Penetration:} To simulate the partial penetration case, we consider both of the above cases as the two extremes and traverse the cases in between with different percentages of CAV inclusion. We adopt a priority-based (yield/stop) traffic movement at the roundabout and on-ramp merging scenarios only for the HDVs, whereas the CAVs are allowed to ignore the traffic signs while exiting a traffic scenario when it is safe to do so.

\subsection{Simulation Results and Analysis}
\begin{figure}[ht!]
    \centering
    \includegraphics[width=1\linewidth]{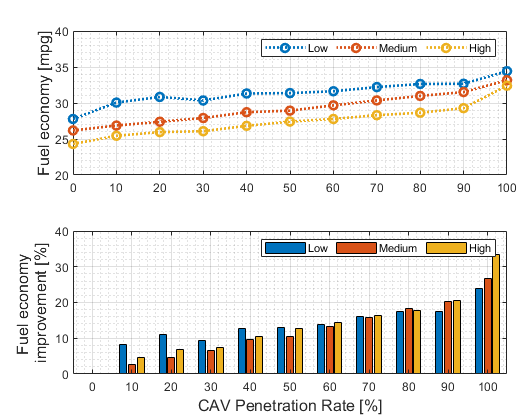}
    \caption{Impact effect of different penetration rates of CAVs on fuel economy.}
    \label{fig:hola}
\end{figure}

\begin{figure}[t!]
    \centering
    \includegraphics[width=1\linewidth]{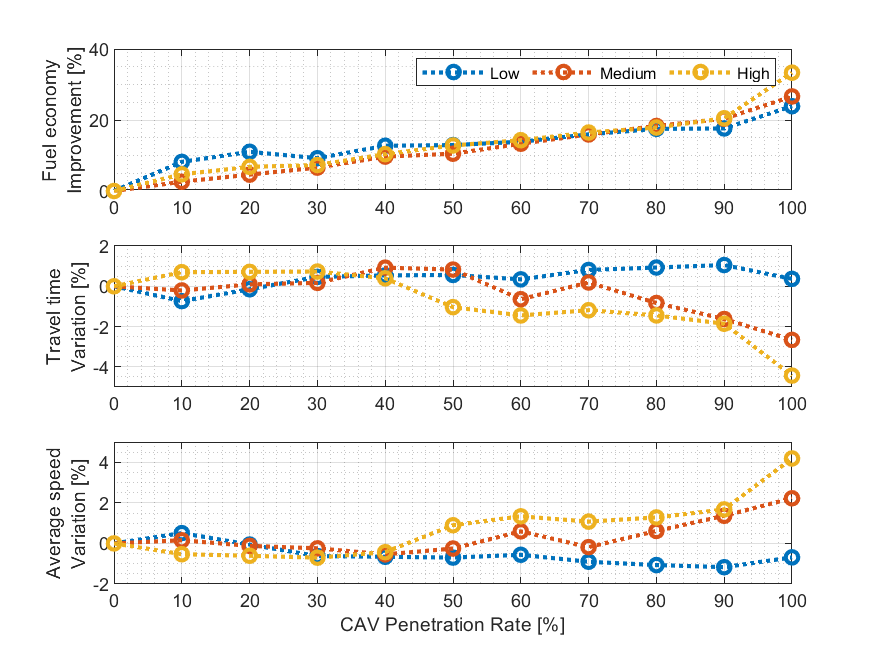}
    \caption{Effect of partial penetration on percentage variation of fuel economy, travel time, and average speed.}
    \label{fig:partpen4}
\end{figure}

We analyzed the implications that 11 different penetration rates of CAVs ranging from 0\% to 100\% may have on fuel economy, travel time improvement as well as mean speed changes. For the fuel consumption analysis, we used the polynomial metamodel presented in \cite{Kamal2013a}.
The simulation results allow several observations. First, for all traffic volumes, fuel economy increases when the penetration rate of CAVs increases, as shown in Fig. \ref{fig:hola}.
Although in the optimal scenario (i.e., 100\% CAVs),  fuel economy improvement is between 24\% and 33\% at all traffic volumes, the improvement for 60\% to 80\% penetration rate is between 13\% to 19\%, with mean and standard deviation as shown in Table \ref{table:mean_impr}. From a general fuel economy improvement point of view, the best CAV penetration rate before the optimal scenario for all the traffic volumes is 70\%, because it has the lowest standard deviation. 
Fuel economy improvement is increasing at all the traffic volumes. Similarly, the average travel time decreases, which represents an improvement in the network. At low traffic volumes, the variations in average speed and travel time, represented by the blue lines in Fig. \ref{fig:partpen4}, are more gradual, in comparison with the other traffic volumes. High traffic volumes, represented by the yellow lines, have the most prominent changes, especially after 50\% penetration. Medium traffic volumes, represented by the red lines, exhibit similar behavior but with some changes around 50\% and 70\% which we discussed more next. 

\begin{table}[h]
\caption{Mean and standard deviation of fuel economy improvement from 60\% to 80\% penetration rate.}
    \begin{tabular}{ |p{3.5cm}|p{1.5cm}|p{1.5cm}| }
\hline
CAV penetration rate[\%] & Mean & Standard Deviation \\
\hline
{60\%} & 13.83\% & 0.41           \\  \hline
{70\%} & 16.12\% & 0.25           \\  \hline
{80\%} & 17.89\% & 0.34          \\  \hline
\end{tabular}
\label{table:mean_impr}
\end{table}

From a general view, the average travel time for high and medium traffic volumes is decreasing while average speed is increasing. In particular, since high traffic volume experiences more stop-and-go driving than the other traffic volumes, the 50\% penetration rate of CAVs has an important impact on travel time decrement and average speed increment, as is going to be shown further in this section. 
At the low traffic volume, the travel time distribution does not improve for 50\% penetration rate, due to the fact that for this traffic volume, CAVs have an impact on fuel economy improvement, but not on average speed, as is shown in Figs. \ref{fig:hola} and \ref{fig:partpen4}. For a 50\% penetration rate in medium traffic volume, average travel time increases. In this situation, most of the vehicles experience an improvement, but some of them get worse times and this affects the average. The same behavior is exhibited for the 70\% penetration rate. 

Considering the information presented in Fig. \ref{fig:partpen4}, the average travel time at the high traffic volume improves for penetration rates above 50\% and has gradual changes for penetrations from 10\% to 40\%. Although Fig. \ref{fig:partpen4} presents a decreasing slope for the high traffic volume, the remarkable improvements occur at 50\% and 100\% penetration rates. At a medium traffic volume, the average travel time is also decreasing, but the behavior is not the same. According to Fig. \ref{fig:partpen4}, the remarkable improvement for this traffic volume occurs at 60\% and maintains a constant improvement after 70\%. Considering the variations of average travel time, and average speed for medium traffic volume at 70\%, the best CAV penetration rate for this volume is 60\%, and not 70\% as was stated just with fuel consumption analysis.

Low traffic volume has its best performance at a 10\% penetration rate, and it remains slightly invariant up to 70\%, where it gets worse. At 100\%, the travel time variation is almost 0\% compared to the baseline, which means that from this point of view, the best CAV penetration rate for low traffic volume is 10\%.

Low traffic volume has almost a constant average travel time. Fig. \ref{fig:partpen4} presents how CAVs penetration improves its fuel economy by 9\% for just 10\% penetration, and it increases as the penetration rate increases. The relationship between fuel economy improvement, travel time variation, and average speed variation is clearly illustrated in Fig. \ref{fig:partpen4}. At all traffic volumes, average speed variation is inversely proportional to travel time improvement, and fuel economy improves as the penetration rate increases. 

\begin{figure}[t!]
    \centering
    \includegraphics[width=1\linewidth]{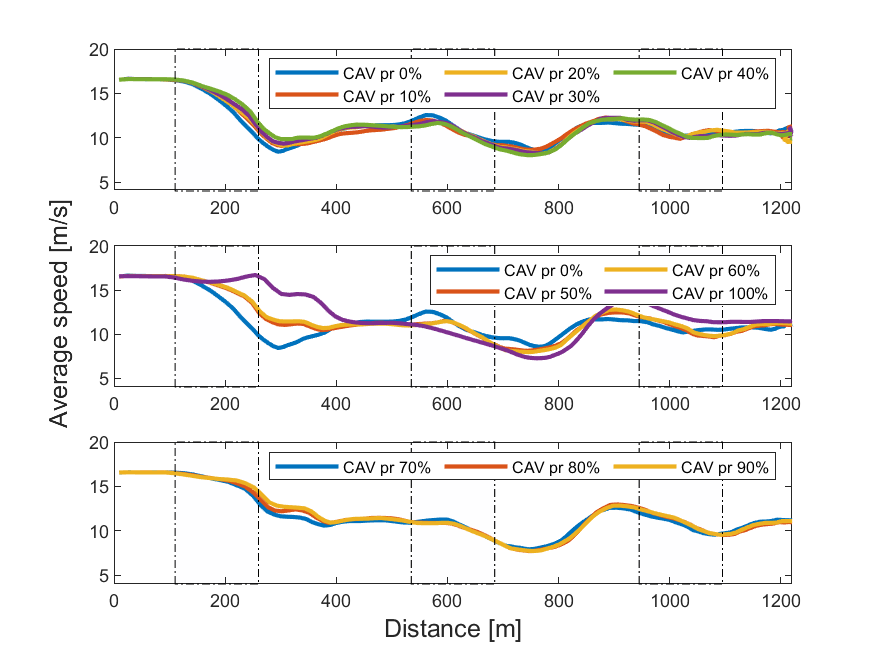}
    \caption{High traffic volume average speed profile for each penetration rate.}
    \label{fig:highavg8}
\end{figure}

The variation of average speed profiles from high traffic volume is shown in Fig. \ref{fig:highavg8}, where the penetrations of interest are compared with the baseline and the optimal scenario. The critical penetrations at this volume are 50\% and 60\%. It is clear how the improvements happen between 200 m and 400 m, which corresponds to a portion of the first traffic scenario (i.e., on-ramp), and the part of the corridor immediately after it, and between 990 m and 1000 m, where the third traffic scenario (i.e., roundabout) occurs. The mean speed has small variations for penetrations smaller than 40\%, and the changes seen for 50\% and 60\% remain very close until 90\% penetration. 

\begin{figure}[t!]
    \centering
    \includegraphics[width=1\linewidth]{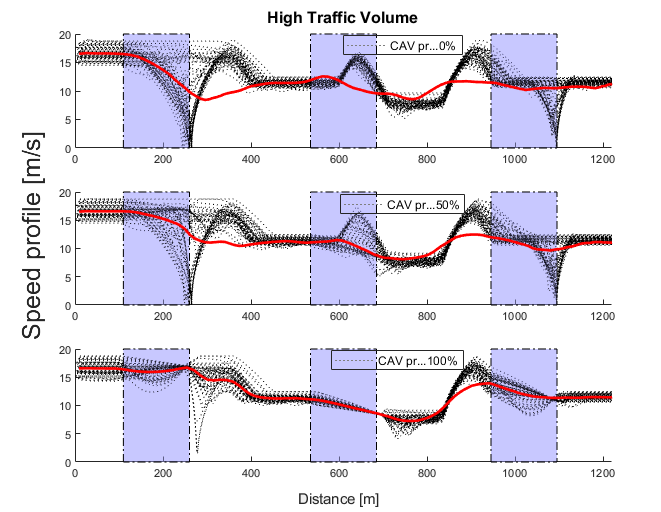}
    \caption{High traffic volume speed profile at the baseline, critical penetration rate, and optimal scenario.}
    \label{fig:speedp8.5}
\end{figure}

Improvements in the speed profile are related to avoiding stop-and-go driving in conflict zones (i.e, the first and third traffic scenarios, and also with speed general reduction in zones like the second traffic scenario. The average speeds at the high traffic volume, shown in Figs. \ref{fig:highavg8} and \ref{fig:speedp8.5}, correspond to the baseline case, 50\% penetration case, and the optimal scenario case. Stop-and-go driving decreases while the penetration rate increases. The speed for the second traffic scenario decreases as well. The average speed in red becomes smoother while the speed profile of all the vehicles tends to converge to a single behavior, as happens in the optimal scenario. 

\subsection{Human-Driven Vehicles vs Connected and Automated Vehicles}
\label{tag:hdvvscav}
\begin{figure}[t!]
    \centering
    \includegraphics[width=1\linewidth]{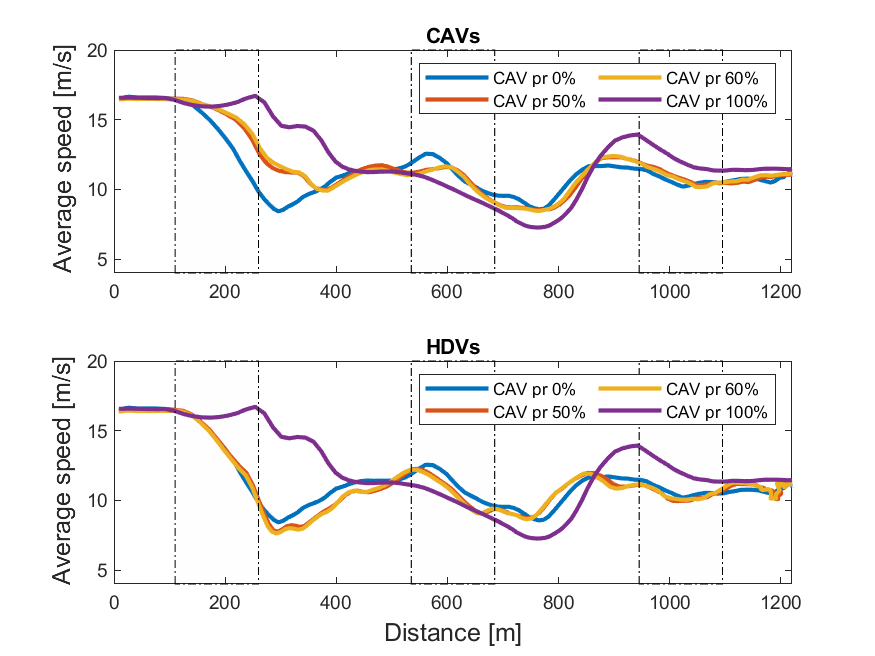}
    \caption{High traffic volume average speed for CAVs and HDVs.}
    \label{fig:hcavhdv9}
\end{figure}

The behavior of HDVs and CAVs differs from each other in mixed traffic scenarios as shown in Fig. \ref{fig:hcavhdv9} at the high traffic volume. The penetration rates of interest are 50\% and 60\%. In both cases, they are compared with the baseline and optimal scenario. Around the rates of interest, it can be seen that the CAVs dominate the speed profile improvement for the first and third traffic scenarios. However, the HDVs have better behavior in the speed reduction zone and a slight improvement near the end of the corridor. The performance improvement of the network for both penetrations of interest is due to the effect of CAVs and HDVs interaction. For a 60\% penetration rate at the medium traffic volume, the HDVs improve the average speed during the transitions before and after the third conflict zone. At the medium traffic volume, the improvement of the network is led by the CAVs. At the low traffic volume, CAVs and HDVs have similar behaviors for all the penetration rates of interest.

\begin{table}[h]
\caption{Fuel economy improvement for HDVs.}
\centering
    \begin{tabular}{|p{1.8cm}|c c c|}
\hline
CAV pr [\%] &  \multicolumn{3}{|c|}{Fuel economy improvement [\%]} \\
 & High & Medium & Low\\
\hline
{40\%} & 1.2\% & -0.7\% & 3.1\%           \\  \hline
{50\%} & 2.7\% & 2\% & 4.1\%           \\  \hline
{60\%} & 2.6\% & 7.4\% & 7.2\%           \\  \hline
{70\%} & 0.5\% & 5.2\% & 9\%           \\  \hline
{80\%} & -1.2\% & 7.5\% & 14.3\%           \\  \hline
{90\%} & 9.8\% & 18.7\% & 23.4\%           \\  \hline
\end{tabular}
\label{table:fuelhdv}
\end{table}

At all three traffic volumes, HDVs exhibit fuel economy improvement as it is shown in Table \ref{table:fuelhdv}. However, the best CAV penetration rate for the HDVs fuel economy is 60\% at all the traffic volumes. From Table \ref{table:fuelhdv} it can be stated that the best fuel economy improvement happens for 90\%. However, for this penetration just 10\% of the vehicles are HDVs, so the high percentage values reported in the table are not giving accurate information about an interesting improvement for HDVs fuel economy in a mixed traffic scenario.


\section{Conclusion} \label{sec:conclusion}
In this paper, we adopted a framework presented in \cite{mahbub2020sae-1} to analyze the implications that different penetration rates of CAVs and their interaction with HDVs can have on fuel economy and travel time. 
The results indicate general improvement performance for all traffic volumes. At low traffic volume, fuel economy improvement is significant, although travel time and average speed do not improve. At medium and high traffic volumes, we observed significant benefits in fuel economy while the penetration rate of CAVs increases. In particular, the improvements in average speed and travel time were observed at 60\% and 50\% penetration rate of CAVs, respectively. 

It is expected that CAVs will gradually penetrate the market, interact with HDVs, and contend communication limitations. 
Ongoing work aims at optimally controlling the CAVs to indirectly control the HDVs and form platoons \cite{mahbub2021_platoonMixed, mahbub2022ACC}. A direction for future research should synergistically integrate human-driving behavior, control theory, and learning in an effort to develop a framework to address a fundamental gap in current methods for the safe co-existence of CAVs with HDVs.

\balance

\bibliographystyle{IEEEtran}
\bibliography{references/ids_lab,references/cav_coord, references/misc}

\begin{thebibliography}{10}
\providecommand{\url}[1]{#1}
\csname url@samestyle\endcsname
\providecommand{\newblock}{\relax}
\providecommand{\bibinfo}[2]{#2}
\providecommand{\BIBentrySTDinterwordspacing}{\spaceskip=0pt\relax}
\providecommand{\BIBentryALTinterwordstretchfactor}{4}
\providecommand{\BIBentryALTinterwordspacing}{\spaceskip=\fontdimen2\font plus
\BIBentryALTinterwordstretchfactor\fontdimen3\font minus
  \fontdimen4\font\relax}
\providecommand{\BIBforeignlanguage}[2]{{%
\expandafter\ifx\csname l@#1\endcsname\relax
\typeout{** WARNING: IEEEtran.bst: No hyphenation pattern has been}%
\typeout{** loaded for the language `#1'. Using the pattern for}%
\typeout{** the default language instead.}%
\else
\language=\csname l@#1\endcsname
\fi
#2}}
\providecommand{\BIBdecl}{\relax}
\BIBdecl

\bibitem{Margiotta2011}
R.~Margiotta and D.~Snyder, ``{An agency guide on how to establish localized
  congestion mitigation programs},'' U.S. Department of Transportation. Federal
  Highway Administration, Tech. Rep., 2011.

\bibitem{chalaki2020TITS}
B.~Chalaki and A.~A. Malikopoulos, ``Time-optimal coordination for connected
  and automated vehicles at adjacent intersections,'' \emph{IEEE Transactions
  on Intelligent Transportation Systems}, pp. 1--16, 2021.

\bibitem{letter2017merging}
C.~Letter and L.~Elefteriadou, ``Efficient control of fully automated connected
  vehicles at freeway merge segments,'' \emph{Transportation Research Part C:
  Emerging Technologies}, vol.~80, pp. 190--205, 2017.

\bibitem{mosebach2016merging}
A.~Mosebach, S.~R{\"o}chner, and J.~Lunze, ``Merging control of cooperative
  vehicles,'' \emph{IFAC-PapersOnLine}, vol.~49, no.~11, pp. 168--174, 2016.

\bibitem{Guanetti2018}
J.~Guanetti, Y.~Kim, and F.~Borrelli, ``{Control of Connected and Automated
  Vehicles: State of the Art and Future Challenges},'' \emph{Annual Reviews in
  Control}, vol.~45, pp. 18--40, 2018.

\bibitem{DeLaFortelle2010}
A.~{de La Fortelle}, ``{Analysis of reservation algorithms for cooperative
  planning at intersections},'' \emph{13th International IEEE Conference on
  Intelligent Transportation Systems}, pp. 445--449, Sep. 2010.

\bibitem{Dresner2008}
K.~Dresner and P.~Stone, ``A multiagent approach to autonomous intersection
  management,'' \emph{Journal of artificial intelligence research}, vol.~31,
  pp. 591--656, 2008.

\bibitem{Ntousakis:2016aa}
I.~A. Ntousakis, I.~K. Nikolos, and M.~Papageorgiou, ``Optimal vehicle
  trajectory planning in the context of cooperative merging on highways,''
  \emph{Transportation Research Part C: Emerging Technologies}, vol.~71, pp.
  464--488, 2016.

\bibitem{Zhao2018ITSC}
L.~Zhao and A.~A. Malikopoulos, ``Decentralized optimal control of connected
  and automated vehicles in a corridor,'' in \emph{2018 21st International
  Conference on Intelligent Transportation Systems (ITSC)}, Nov 2018, pp.
  1252--1257.

\bibitem{mahbub2020decentralized}
A.~I. Mahbub, A.~A. Malikopoulos, and L.~Zhao, ``Decentralized optimal
  coordination of connected and automated vehicles for multiple traffic
  scenarios,'' \emph{Automatica}, vol. 117, no. 108958, 2020.

\bibitem{mahbub2020ACC-2}
A.~M.~I. Mahbub, A.~A. Malikopoulos, and L.~Zhao, ``Impact of connected and
  automated vehicles in a corridor,'' in \emph{Proceedings of 2020 American
  Control Conference, 2020}.\hskip 1em plus 0.5em minus 0.4em\relax IEEE, 2020,
  pp. 1185--1190.

\bibitem{kotsialos2004Merging}
A.~Kotsialos and M.~Papageorgiou, ``Nonlinear optimal control applied to
  coordinated ramp metering,'' \emph{IEEE Transactions on control systems
  technology}, vol.~12, no.~6, pp. 920--933, 2004.

\bibitem{Kumaravel:2021wi}
S.~Kumaravel, A.~A. Malikopoulos, and R.~Ayyagari, ``Optimal coordination of
  platoons of connected and automated vehicles at signal-free intersections,''
  \emph{IEEE Transactions on Intelligent Vehicles}, pp. 1--1, 2021.

\bibitem{alessandrini2015automatedmixed2060}
A.~Alessandrini, A.~Campagna, P.~Delle~Site, F.~Filippi, and L.~Persia,
  ``Automated vehicles and the rethinking of mobility and cities,''
  \emph{Transportation Research Procedia}, vol.~5, pp. 145--160, 2015.

\bibitem{zheng2017platooning}
Y.~Zheng, S.~E. Li, K.~Li, and W.~Ren, ``Platooning of connected vehicles with
  undirected topologies: Robustness analysis and distributed h-infinity
  controller synthesis,'' \emph{IEEE Transactions on Intelligent Transportation
  Systems}, vol.~19, no.~5, pp. 1353--1364, 2017.

\bibitem{sharon2017protocol}
G.~Sharon and P.~Stone, ``A protocol for mixed autonomous and human-operated
  vehicles at intersections,'' in \emph{International Conference on Autonomous
  Agents and Multiagent Systems}.\hskip 1em plus 0.5em minus 0.4em\relax
  Springer, 2017, pp. 151--167.

\bibitem{milanes2013cooperative}
V.~Milan{\'e}s, S.~E. Shladover, J.~Spring, C.~Nowakowski, H.~Kawazoe, and
  M.~Nakamura, ``Cooperative adaptive cruise control in real traffic
  situations,'' \emph{IEEE Transactions on intelligent transportation systems},
  vol.~15, no.~1, pp. 296--305, 2013.

\bibitem{jin2017fuel_mixplatoon}
I.~G. Jin, G.~Orosz, D.~Hajdu, T.~Insperger, and J.~Moehlis, ``To delay or not
  to delay—stability of connected cruise control,'' in \emph{Time Delay
  Systems}.\hskip 1em plus 0.5em minus 0.4em\relax Springer, 2017, pp.
  263--282.

\bibitem{hajdu2019robust}
D.~Hajdu, I.~G. Jin, T.~Insperger, and G.~Orosz, ``Robust design of connected
  cruise control among human-driven vehicles,'' \emph{IEEE Transactions on
  Intelligent Transportation Systems}, vol.~21, no.~2, pp. 749--761, 2019.

\bibitem{orosz2016connected}
G.~Orosz, ``Connected cruise control: modelling, delay effects, and nonlinear
  behaviour,'' \emph{Vehicle System Dynamics}, vol.~54, no.~8, pp. 1147--1176,
  2016.

\bibitem{guler2014_Intersection}
S.~I. Guler, M.~Menendez, and L.~Meier, ``Using connected vehicle technology to
  improve the efficiency of intersections,'' \emph{Transportation Research Part
  C: Emerging Technologies}, vol.~46, pp. 121--131, 2014.

\bibitem{Gipps1981}
P.~Gipps, ``{A behavioural car-following model for computer simulation},''
  \emph{Transportation Research Part B: Methodological}, vol.~15, no.~2, pp.
  105--111, 1981.

\bibitem{wiedemann1974}
R.~Wiedemann, ``Simulation des strassenverkehrsflusses.'' 1974.

\bibitem{zhang2018penetration}
Y.~Zhang and C.~G. Cassandras, ``The penetration effect of connected automated
  vehicles in urban traffic: an energy impact study,'' in \emph{2018 ieee
  conference on control technology and applications (ccta)}.\hskip 1em plus
  0.5em minus 0.4em\relax IEEE, 2018, pp. 620--625.

\bibitem{ding2020penetration}
J.~Ding, H.~Peng, Y.~Zhang, and L.~Li, ``Penetration effect of connected and
  automated vehicles on cooperative on-ramp merging,'' \emph{IET Intelligent
  Transport Systems}, vol.~14, no.~1, pp. 56--64, 2020.

\bibitem{wan2016optimalmixed}
N.~Wan, A.~Vahidi, and A.~Luckow, ``Optimal speed advisory for connected
  vehicles in arterial roads and the impact on mixed traffic,''
  \emph{Transportation Research Part C: Emerging Technologies}, vol.~69, pp.
  548--563, 2016.

\bibitem{kreidieh2018dissipating}
A.~R. Kreidieh, C.~Wu, and A.~M. Bayen, ``Dissipating stop-and-go waves in
  closed and open networks via deep reinforcement learning,'' in \emph{2018
  21st International Conference on Intelligent Transportation Systems
  (ITSC)}.\hskip 1em plus 0.5em minus 0.4em\relax IEEE, 2018, pp. 1475--1480.

\bibitem{wu2017framework}
C.~Wu, K.~Parvate, N.~Kheterpal, L.~Dickstein, A.~Mehta, E.~Vinitsky, and A.~M.
  Bayen, ``Framework for control and deep reinforcement learning in traffic,''
  in \emph{2017 IEEE 20th International Conference on Intelligent
  Transportation Systems (ITSC)}.\hskip 1em plus 0.5em minus 0.4em\relax IEEE,
  2017, pp. 1--8.

\bibitem{mahbub2020sae-1}
A.~M.~I. Mahbub, V.~Karri, D.~Parikh, S.~Jade, and A.~A. Malikopoulos, ``A
  decentralized time- and energy-optimal control framework for connected
  automated vehicles: From simulation to field test,'' in \emph{SAE Technical
  Paper 2020-01-0579}.\hskip 1em plus 0.5em minus 0.4em\relax SAE
  International, 2020.

\bibitem{bryson1975applied}
A.~E. Bryson and Y.~C. Ho, \emph{Applied optimal control: optimization,
  estimation and control}.\hskip 1em plus 0.5em minus 0.4em\relax CRC Press,
  1975.

\bibitem{mahbub2020Automatica-2}
A.~M.~I. Mahbub and A.~A. Malikopoulos, ``{Conditions to Provable System-Wide
  Optimal Coordination of Connected and Automated Vehicles},''
  \emph{Automatica}, vol. 131, no. 109751, 2021.

\bibitem{ptv2014ptv}
P.~Group \emph{et~al.}, ``Ptv vissim 7 user manual,'' \emph{Germany: PTV
  GROUP}, 2014.

\bibitem{Vissim_microscopic}
M.~Fellendorf and P.~Vortisch, ``Microscopic traffic flow simulator vissim,''
  in \emph{Fundamentals of traffic simulation}.\hskip 1em plus 0.5em minus
  0.4em\relax Springer, 2010, pp. 63--93.

\bibitem{Kamal2013a}
M.~Kamal, M.~Mukai, J.~Murata, and T.~Kawabe, ``{Model Predictive Control of
  Vehicles on Urban Roads for Improved Fuel Economy},'' \emph{IEEE Transactions
  on Control Systems Technology}, vol.~21, no.~3, pp. 831--841, 2013.

\bibitem{mahbub2021_platoonMixed}
A.~M.~I. Mahbub and A.~A. Malikopoulos, ``{A Platoon Formation Framework in a
  Mixed Traffic Environment},'' \emph{IEEE Control Systems Letters (LCSS)},
  vol.~6, pp. 1370--1375, 2021.

\bibitem{mahbub2022ACC}
A.~M. {Ishtiaque Mahbub} and A.~A. {Malikopoulos}, ``{Platoon Formation in a
  Mixed Traffic Environment: A Model-Agnostic Optimal Control Approach},''
  \emph{Proceedings of 2022 American Control Conference}, Oct. 2022 (to
  appear).

\end{thebibliography}

\end{document}